\documentclass[prl,aps,twocolumn,showpacs,floatfix]{revtex4}

\usepackage{graphicx}
\usepackage{amssymb}
\usepackage{amsmath}
\usepackage{amsfonts}
\usepackage{epstopdf}
\usepackage{verbatim}
\usepackage[normalem]{ulem}

\begin{document}

\title{Structural distortion behind the nematic superconductivity in Sr$_x$Bi$_2$Se$_3$.}

\author{ A.\,Yu.~Kuntsevich$^{a}$\thanks {e-mail:
alexkun@lebedev.ru}, M.A. Bryzgalov$^{b}$, V.A. Prudkoglyad$^{a}$, V.P. Martovitskii$^{a}$, Yu.G. Selivanov$^{a}$, E.G. Chizhevskii$^{a}$}

\affiliation{$^a$P.\,N.~Lebedev Physics Institute, Moscow 119991, Russia}
\affiliation{$^b$Moscow Institute of Physics and Technology, Dolgoprudny, Moscow region 141700, Russia}

\begin{abstract}
An archetypical layered topological insulator Bi$_2$Se$_3$ becomes superconductive upon doping with Sr, Nb or Cu. Superconducting properties of these materials in the presence of in-plane magnetic field demonstrate spontaneous symmetry breaking: 180$^\circ$-rotation symmetry of superconductivity versus 120$^\circ$-rotation symmetry of the crystal. Such behavior brilliantly confirms nematic topological superconductivity. To what extent this nematicity is due to superconducting pairing in these materials, rather than due to crystal structure distortions? This question remained unanswered, because so far no visible deviations from the 3-fold crystal symmetry were resolved in these materials.
To address this question we grow high quality single crystals of Sr$_x$Bi$_2$Se$_3$, perform detailed  X-ray diffraction and magnetotransport studies and reveal that the observed superconducting nematicity direction correlates with the direction of small structural distortions in these samples( $\sim 0.02$\% elongation in one crystallographic direction). Additional anisotropy comes from orientation of the crystallite axes. 2-fold symmetry of magnetoresistance observed in the most uniform crystals well above critical temperature demonstrates that these structural distortions are nevertheless strong enough. Our data in combination with strong sample-to-sample variation of the superconductive anisotropy parameter are indicative for significance of the structural factor in the apparent nematic superconductivity in Sr$_x$Bi$_2$Se$_3$.\end{abstract}

\pacs{72.15.Gd, 74.25.Fy, 74.62.Bf}
\maketitle

\section{Introduction}

The most studied topological insulator material, Bi$_2$Se$_3$, becomes superconductive being doped with Sr, Nb or Cu, with $T_c$ around 3K and $H_{c2}$ about a few Tesla\cite{radioactive,pressure,NMR,HeatCap,deVisser,andostm,chinese,asaba,jacs,srbisenew,stm,shen,shruti, srbises, levy, kobayashiNb,intstress, prmat, neupane}. Nature of this superconductivity (topological or not) is in the focus of discussion during the last few years. On the theoretical site, strong spin-orbit interaction in the parent compound may favor triplet topological superconductivity\cite{fu}, moreover this superconductivity is expected to be insensitive to disorder\cite{michaeli}, inherent to layered chalcogenides. A fingerprint of topological superconductivity would be Majorana fermion, showing itself up as a zero-bias feature in I-V-characteristic. First soft-contact measurements in superconducting Cu$_x$Bi$_2$Se$_3$ \cite{sasaki} have shown a promising zero-bias peak. However, later tunnel spectroscopy measurements\cite{levy} have clearly shown an s-wave pairing without any in-gap states.  Recently this superconductivity(SC) was found to be nematic, i.e. superconducting properties depend strongly on the in-plane orientation of the magnetic field \cite{NMR, asaba, HeatCap, deVisser, andostm, shen,chinese, srbisenew}. Critical magnetic fields, magnetization, resistivity, specific heat, and Knight shift have 180$^\circ$ in-plane rotation symmetry, contrary to the trigonal (120$^\circ$) crystal symmetry. An explanation for such nematicity was again suggested within the topological superconductivity model with two component order parameter\cite{fu2,venderbos}, and most of the data brilliantly confirm this theoretical approach. Very recently nematicity of superconducting properties (distortion of Abrikosov vortices)in combination with the zero bias peak at the vortex cores were directly visualized using more thorough scanning tunneling spectroscopy\cite{andostm}.

Nematicity of both thermodynamic and transport properties in these materials was found to be linked to crystallographical directions\cite{NMR,HeatCap,srbisenew,deVisser,chinese,andostm}. Apparently, real crystals never grow with perfect 3-fold symmetry, there should be some structural distortion. If this distortion is tiny, it will not affect the transport above $T_c$ and will not couple to superconducting Hamiltonian just driving the orientation of the superconducting nematicity. In case of stronger deviations from perfect R$\bar{3}$m symmetry, both superconductivty and transport properties above $T_c$ will inherit this asymmetry. In the limit of strong distortion, the nematicity is predominantly determined by it. In general, there is a question to what extent the nematicty is intrinsic(i.e. due to superconducting pairing mechanism) or emergent (i.e. driven by this structural distortion).

It should be noted that studies of the structural imperfections require samples with the highest crystalline quality. In a number of papers \cite{deVisser, chinese, srbisenew, jacs, pressure, stm, shruti, prmat} Sr$_x$Bi$_2$Se$_3$ crystals were prepared with an ordinary melt-growth technique without a certain growth direction. A few experiments used samples made by Bridgman method \cite{neupane}. In most of the present experimental papers only powder X-ray diffraction(XRD)\cite{shruti,jacs} and Laue\cite{deVisser,chinese,andostm} patterns are presented. Absence of high resolution single crystal X-ray diffraction data is an indirect indicator that the crystal quality was not ultimately optimized. Indeed, Cu$_x$Bi$_2$Se$_3$ systematically demonstrates rather low superconducting fraction \cite{sasaki,NMR}. First Sr$_x$Bi$_2$Se$_3$ \cite{shruti} and Nb$_x$Bi$_2$Se$_3$\cite{kobayashiNb} demonstrated admixture of the second phases. The block structure of the grown crystals was almost not discussed at all.  Only recently, XRD-data supported high enough structural quality Sr$_x$Bi$_2$Se$_3$ \cite{srbisenew} and Nb$_x$Bi$_2$Se$_3$\cite{radioactive,shen} crystals were reported with $\sim$100\% shielding fraction.

 To achieve high crystallinity we grew Sr$_x$Bi$_2$Se$_3$ using a Bridgman method with post growth annealing, which is expected to produce single crystals consisting of blocks all aligned along the growth direction.  The obtained crystals were thoroughly studied with X-Ray diffraction and magnetotransport both below and above $T_c$. We do find that magnetoresistance well above T$_c$ has the same two-fold in-plane asymmetry as superconductivity does, in agreement with slight triclinic distortion of the lattice found in XRD studies. More interestingly, in samples big enough(containing more than one block) we reveal that the structural anisotropy is also a consequence of the block alignment.
    Our results give arguments in favor of strong coupling between structural distortion and superconductivity, thus raising a question to what extent the nature of nematicity is topological.

\begin{figure*}
\includegraphics[width=16cm]{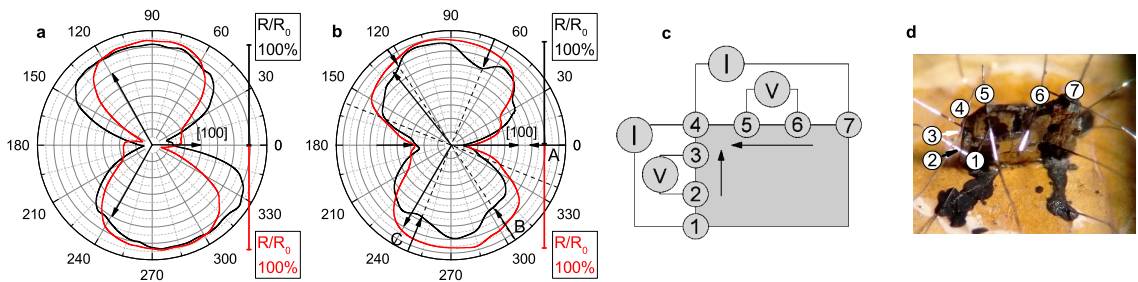}
\caption{Anisotropy of superconductivity.(a) Anisotropic magnetoresistance $R_{7456}$ (red line) and $R_{1423}$ (black line) in small Sr$_{0.1}$Bi$_2$Se$_3$ \#306-1 sample at $T$ =2.3K and in-plane field $B$ = 0.8T. (b) Same for big Sr$_{0.1}$Bi$_2$Se$_3$ \#306-2 sample at in-plane field $B$ = 0.4T. Inbound arrows indicate the directions A, B, and C of minima in magnetoresistance due to superconductivity in various crystallites. Outbound arrows indicate crystallographic directions. (c) Schematics of the sample. (d) Photo of the sample Sr$_{0.1}$Bi$_2$Se$_3$ \#306-2 with enumerated contacts.}
\label{fig-amr-1}
\end{figure*}

\begin{figure}
\includegraphics[width=9cm]{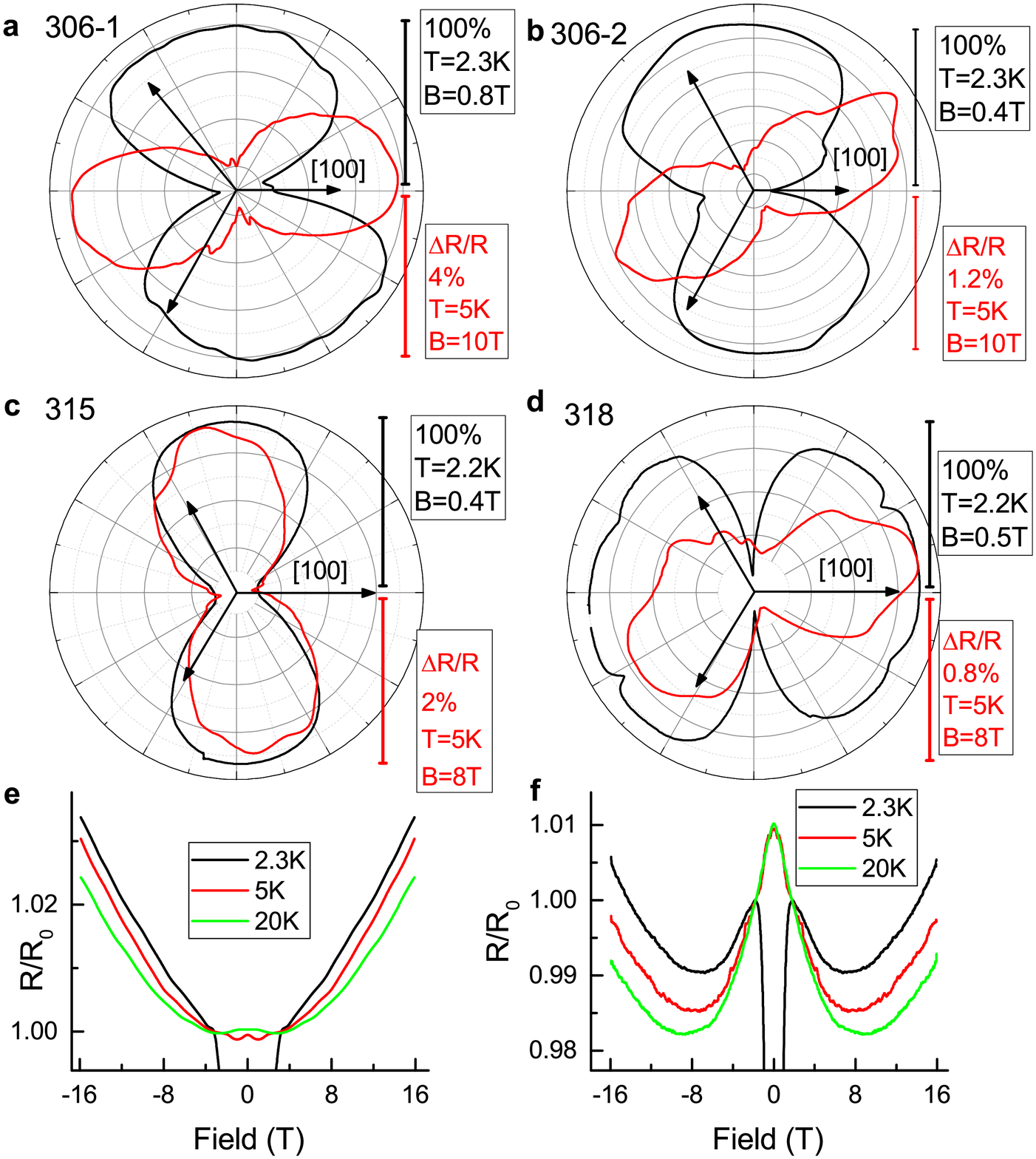}
\caption{ (a-d) Anisotropy of superconductivity(red) versus anisotropy of normal magnetoresistance(black) $(R-R_0)/R_0$ for various samples. Sample number, magnetic field, temperatures and the scale of the effect are indicated in the panels. The outbound arrows indicate the in plane crystallographic directions and the shortest (longest) arrow shows schematically the direction of lattice (compression)elongation, determined from XRD studies. (e),(f) Field dependence of the magnetoresistance at different temperatures for directions of maximum $H_{c2}$ and minimum $H_{c2}$ accordingly for sample 306-2. Current flow direction corresponds to the direction of maximum $H_{c2}$. The temperatures are indicated in the panels.}
\label{fig-amr-2}
\end{figure}

\section{Results}

\subsection{Samples}

A series of Sr$_x$Bi$_2$Se$_3$ samples with nominal Sr content of $x = 0.10$,  0.15 and 0.20 were prepared using modified Bridgman method \cite{aleshenko}. High purity elemental Bi,  Se (99.999\%) and Sr (99.95\%)  in the desired molar ratio were loaded in quartz ampoules inside inert atmosphere glove box.  Sealed evacuated tubes were heated at 850 $^\circ$C for 24 hours with periodic stirring followed by a slow cooling to 620 $^\circ$C  at a rate of $\sim$2 $^\circ$C per hour. The samples were then annealed at 620 $^\circ$C  for 48 hour and water quenched. The crystals obtained by this method had a mirrorlike surface and were readily cleaved along the basal plane. Their structure was characterized by single crystal XRD (Panalytical XPert Pro MRD Extended). High-resolution XRD studies gave  reproducible results, since they were performed in air-conditioned room at 22.5$\pm$0.5 $^\circ$C and could not be affected by temperature drifts.

 Studies of the crystals grown started with the X-ray diffraction selection of the proper samples. Sr$_x$Bi$_2$Se$_3$ single crystals obtained, as our XRD-studies show, always consisted of blocks (crystallites) with lateral dimensions $0.05-0.5$ mm (evaluated from typical number of peaks at the rocking curves). The blocks had the same structure, slight variation of the c-axis lattice parameter and misorientation smaller than 1$^{\circ}$ with respect to each other. For further transport measurement and detailed structural investigations we cleaved small pieces of crystals containing only one or a few blocks. The lateral dimensions of these samples were in the range 0.6-3 mm and thickness was in the range 50-150 $\mu$m.

 All crystals, selected for transport measurements, had a dominating block and all subsequent detailed studies of the structural distortions ($2\theta/\omega$ scanning) were performed on this dominating block within the cleaved sample. Moreover, we expect that it is a dominating block that governs transport properties.

We should note here, that (i)block structure is the intrinsic property of these materials (see Appendix 1), (ii)our crystals are of the highest quality(see Appendix 3), and (iii)the observation of the discovered effects become possible only due this high quality and a feedback between XRD and transport measurements.

\subsection{Transport measurements}
For transport studies single crystals (4-wire resistance 10-100 mOhm) were mounted on the holders and the contact wires (diameter 0.02 mm) were glued with either silver or graphite paint (size of the paint drop was about 0.1 mm and resistance about 20-100Ohm per contact). We aligned the basal plane and plane of rotation by eye with precision $\sim3^o$ parallel to each other.
Magnetotransport measurements were performed with four-terminal scheme using lock-in detection at frequencies 13-190 Hz and measurement currents up to 500 uA. We have checked that measurement current did not overheat the samples(i.e. did not shift the SC transition temperature). We used Cryogenics 21T/0.3K, dry CFMS 16T, and Quantum Design PPMS 9T systems equipped with platforms, that allowed to rotate the samples in-situ to $\sim$360$^{\circ}$. Magnetic field sweeps were performed from positive to negative direction and magnetoresistance was obtained by symmetrization of the data. For measurements of in-plane angular dependence of magnetoresistance (AMR), similarly to the previous investigators\cite{NMR, HeatCap, chinese, deVisser, asaba, srbisenew,shen}, we applied constant magnetic field at fixed temperature and rotated the sample. To ensure that the AMR patterns are not affected by temperature drift, the temperature was stabilized with better than 0.1K precision.  To exclude inevitable admixture of the Hall effect we subtracted the lowest harmonic $A\cos (\varphi) + B\sin (\varphi)$ from the AMR. The arguments why such subtraction is needed are summarized in the Appendix 4.

 None of the studied crystals demonstrated perfect uniformity of superconducting properties, i.e. anisotropic magnetoresistance (AMR) depends on choice of potential probes and usually deviated from an ideal 8-like shape.  For the single-block samples the highest uniformity is expected. Fig. \ref{fig-amr-1}a shows AMR for sample Sr$_{0.1}$Bi$_2$Se$_3$ \#306-2 (1x1x0.05 mm) for two current flow directions, indicated in panel \ref{fig-amr-1}c by arrows. One can see that the nematicity is almost insensitive to the current flow direction. This observation is in-line with previous magnetotransport studies\cite{chinese} where current flow was perpendicular to basal plane as well as with thermodynamic studies ( specific heat\cite{HeatCap} and magnetometry\cite{asaba}), and it was also recently confirmed in Ref.\cite{srbisenew} . Interestingly, in all our samples, strongest SC direction (where $H_{c2}$ is maximal) either roughly coincides or is perpendicular to crystallographical axis in the basal plane, indicated by arrows in Figs.\ref{fig-amr-1},\ref{fig-amr-2},\ref{fig-xrd-2}. Exactly the same alignment was observed earlier in Ref.\cite{chinese}. Thus, we have systematically reproduced all previous results\cite{deVisser, chinese,srbisenew} of anisotropic in-plane magnetotransport in our small samples.  The superconducting properties, including the two-fold axis direction, are stable i.e. don't change after two months exposure at ambient conditions and insensitive to thermal cycling.

 AMR for large single crystal \#306-2 (2x3x0.05 mm), shown in Fig.\ref{fig-amr-1}b, suggests the presence of several SC domains. Indeed, for current flow direction from contact 7 to contact 4 (red curve), AMR is similar to small samples, whereas for perpendicular current flow (from 1 to 4, black curve) besides the main SC direction (A), two other directions emerge (B and C). In the particular case of Sample \#306-1, the angle between axes A, B and C is 60$^\circ$. This observation clearly evidences for presence of domains with various orientation of SC axis located close to contacts 1, and 2. SC axis in these domains is aligned along different crystallographical directions than in the main domain. 
 Thus, we show for the first time the multidomain character of the single crystalline superconducting Sr$_x$Bi$_2$Se$_3$ samples. We believe that different SC domains correspond to different crystallites.  Even more dramatic example of SC multidomain structure and block structure is discussied in Appendix 2. In-plane manifold structure was also observed in Nb$_x$Bi$_2$Se$_3$ with magnetization measurements\cite{asaba}. The structure was unambiguously interpreted as fingerprint of intermixing between SC spontaneous symmetry breaking and crystalline 3-fold symmetry. Our data pose a question whether this pattern is simply a consequence of multidomain sample structure.

Absence of the structural indications of the three-fold symmetry breaking in previous works\cite{NMR, HeatCap, deVisser, chinese, asaba,srbisenew, shen},  and guidance by the Curie principle that the symmetry should be lowered\cite{Curie}  motivated us to perform detailed studies of magnetoresistance above $T_c$.
{Apart from previous investigators \cite{deVisser, chinese, srbisenew} we (i)performed measurements on the crystals with a dominating block (ii) extended the range of magnetic fields to 10 Tesla in order to expand possible assymetries and (iii) monitored tiny  features of the magnetoresistance.

 Magnetoresistance $\rho(B)$ in these system appeared to be non-monotonic and quite complex (See Fig. \ref{fig-amr-2} e,f). In particular, in the low magnetic field regime, a negative magnetoresistance emerges, that is non-trivial for metallic system. This magnetoresistance is discussed in Appendix 5. In most of our samples we managed to find apparent 2-fold symmetries.
Fig.\ref{fig-amr-2}a shows an angular dependence of the SC suppression (black curve) at $T$ = 2.3K and magnetoresistance (red curve) of the same sample at $T$ = 5K $B$ = 10T. The direction of the strongest SC for this sample coincides with the direction of maximal magnetoresistance and with one of the in-plane crystalline directions. For the in-plane 3-fold symmetric system, current flow direction should generate asymmetry in the system and if it was the only case, magnetoresistance would depend on angle between current flow and magnetic field.  However, experimentally applying current flow in perpendicular direction doesn't rotate AMR, but rather changes it in complex manner: $A\cdot cos(2\varphi)$ component becomes suppressed. Such a change is an indicator of the crystalline anisotropy, and it was observed systematically in various samples.
 Moreover, as we rotate current flow direction out of the optimal path, the first harmonic ($\propto \cos(\varphi)$), responsible for the Hall effect due to deviation of the current flow from the parallel planes, significantly increases. This fact indirectly indicates that transport current bypasses some obstacles and flows in z-direction.

 While in some of our samples (Fig.\ref{fig-amr-2}a,b) direction of maximal magnetoresistance coincides with direction of minimal $H_{c2}$, in the others they are roughly perpendicular (Fig.\ref{fig-amr-2}c,d). This is not unusual, because the observation, that in Sr$_x$Bi$_2$Se$_3$ a certain crystallographic axis can be either parallel or perpendicular to the nematicity direction was already reported in Ref.\cite{chinese}.}

Comparing the single domain (Fig.\ref{fig-amr-2}a ) and multidomain (Fig. \ref{fig-amr-2}b), pieces of the same \#306 crystal, we find that the magnitude of the magnetoresistance in the single domain sample is larger than in the multiblock one. It means that anisotropy in magnetoresistance above $T_c$ is governed by structural distortion, rather than by grain boundaries. In multidomain sample \#306-2 admixtures of domains with different orientations (demonstrated by Fig.\ref{fig-amr-1}b) apparently weakens the magnetoresistance. This fact is a strong indication that the same structural distortion governs the nematic magnetoresistance in both SC and normal state.

The anisotropic magnetoresistance in the normal state slowly weakens with temperature (Fig. \ref{fig-amr-2}e,f), and preserves its 2-fold symmetry up to 200K. In other words, the SC in-plane axis can be anticipated from AMR not only close to $T_c$, as recently predicted in Ref.\cite{nematheory}, but also well above $T_c$, i.e, this effect is of structural nature.

\begin{figure*}
\includegraphics[width=16cm]{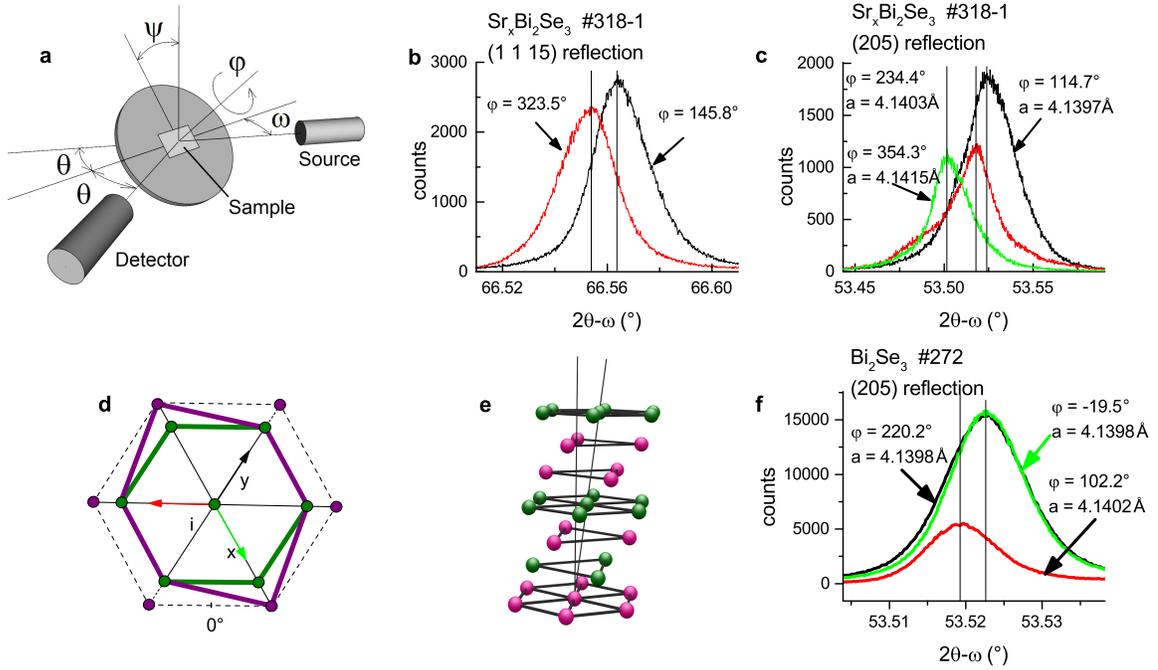}
\caption{X-ray diffraction studies of the structural distortions in the sample \#318-1. (a) Schematics of XRD measurement. $(2\theta-\omega)$-scanning curves with triple crystal-analyzer in the grazing diffraction geometry on the 3-fold reflection (2 0 5), $\psi = 72.6^\circ$ (b); and on the 6-fold reflection (1 1 15), $\psi = 42.7^\circ$ (c) for various $\varphi$ for Sr$_{0.2}$Bi$_2$Se$_3$ sample \#318-1. (d) In-plane one-axis elongation structural distortion. Green points correspond to undoped Bi$_2$Se$_3$. Purple outer points correspond to expected Sr-induced lattice expansion. Purple hexagon is a schematic representation of the lattice distortion corresponding to XRD patterns from panel (b). (e) c-axis inclination structural distortion. (f) $(2\theta-\omega)$-scanning curves on the (2 0 5) reflection (similar to panel (c)) for bare Bi$_2$Se$_3$ \#272 sample demonstrating an order of magnitude smaller structural distortion than the Sr-doped crystals).}
\label{fig-xrd}
\end{figure*}

\subsection{Structural studies: triclinic distortions}

 In the most of previous structural studies of doped SC Bi$_2$Se$_3$ \cite{jacs,deVisser,chinese,srbisenew, andostm} either Laue or the powder diffraction was used, that didn't allow to resolve fine structural imperfections.  It should be noted, that for our high-resolution structural measurements, high crystalline quality (large sizes of uniform blocks, absence of bending) is needed and special precautions were taken in order to avoid systematic shifts of reflection maximum. Before recording each (2$\theta - \omega$) curve the sample was (i) rotated around the $\psi$ axis to achieve vertical position for the diffraction plane and (ii) shifted along the goniometer X- and Y- directions in order to achieve illumination of the whole sample by X-Ray. The precision of the lattice parameter measurements from the (2$\theta /\omega$) curves in our case was limited by crystal quality and was better than 0.0001 {\AA}(see book \cite{xrdbook} for reference on XRD measurement techniques). We use reflection indices in the hexagonal lattice notations with omitted triple index in the basal plane, which is equal to minus sum of the first two indices. Due to non-ideal planar geometry of the samples,
the $\varphi$ rotation axis of the diffractometer sometimes is not exactly parallel to the crystallographic $c$-axis of the studied block. So, despite tri-fold and 6-fold symmetry of the (205) and (1 1 15)  reflections respectively, the angular distance $\Delta\varphi$ between reflections of these families might be not an exact integer of 120$^\circ$/60$^\circ$. Note, that due to  above described sample adjustment procedure  this misalignment does not affect high-precisson $2\theta /\omega$ angle measurements, and, as a consequence, precise lattice constant determination.

Strong (0 0 n) symmetrical reflections, explored previously for Sr$_x$Bi$_2$Se$_3$ by single crystal XRD in Refs.\cite{jacs,shruti}, are sensitive only to the lattice parameter value along the c-axis and do not allow to study in-plane lattice anisotropy.
 In order to find the lattice distortion in the basal plane, we used intensive (2 0 5) and (1 1 15) asymmetrical reflections, resolved in grazing diffraction geometry (Fig.~\ref{fig-xrd}a), because the diffraction angle values for these reflections are smaller than the inclination angles to the basal plane ($\theta_{B(2\,0\,5)}=26.75^\circ, \psi_{(2\,0\,5)/(0\,0\,1)}=72.62^\circ$;  $\theta_{B(1\,1\,15)}=33.27^\circ, \psi_{(1\,1\,15)/(0\,0\,1)}=42.7^\circ$). $\varphi $-scanning curves of the (2 0 5) reflection apparently demonstrate three-peak 120$^\circ$ rotation symmetry of the studied crystalline structure.  This reflection is already sensitive mainly to lattice parameter in the basal plane.

In order to resolve the small structural distortion effects we used a detector with the triple crystal-analyzer 3$\times$Ge(220) for high resolution  $(2\theta-\omega)$-scanning curves  for each of these 3-fold (2 0 5) and 6-fold (1 1 15) peaks.  It is evident that when lattice distortions are absent, the maxima of (2 0 5) and (1 1 15) reflections in all azimuthal positions must be unchanged. However, as we have systematically observed in all our samples, the positions of these peaks change upon in-plane rotation. Fig.\ref{fig-xrd}b shows  an example for the most anisotropic sample Sr$_{0.2}$Bi$_2$Se$_3$ \#318-1 (anisotropy factor ${H_{c2}}^{max}/{H_{c2}}^{min} = 8$). Variation of the (2 0 5) reflection peak position for $\varphi = 114.7^\circ$ and $\varphi = 354.3^\circ$ is about 0.02$^\circ$ in Fig. \ref{fig-xrd}b, and corresponds to 0.02\% lattice parameter elongation along $a$-axis, as shown schematically in Fig.\ref{fig-xrd}d. We observed either alongation (samples \#315, \#318, and \#308) or shortening (\#306) of the lattice parameter along one of axis with respect to the others. The arrows in Fig. \ref{fig-amr-2}a-d indicate the directions of the crystalline axes, and the shortest (or longest, depending on a sample) is shown by length. In all cases the direction of the SC axis was either parallel or perpendicular to in-plane direction of the maximal crystalline deformation, similarly to Ref.~\cite{chinese}.
Concerning the normal state magnetoresistance, in \#306 sample (the shortest axis, Fig.~\ref{fig-amr-2}a-b ) the two-fold magnetoresistance below an above $T_c$ are perpendicular, while in samples \#315 and \#318 (the longest axis, Figs. Fig.\ref{fig-amr-2}c and Fig.\ref{fig-amr-2}d, respectively) they are parallel. This fact indicates correlation between tensile or compressive structural distortion and nematicity of the electronic properties.

The same diffraction patterns as in Fig.\ref{fig-xrd}b might be caused also by a small deviation of the c-axis from the perpendicular to the basal plane(Fig.\ref{fig-xrd}e). In order to evaluate the role of this second type of distortion we used the (1 1 15) reflection that nominally has six-fold rotational symmetry. Were the c-axis not inclined, the positions of (1 1 15) and (-1 -1 15) reflection maxima obtained after 180$^\circ$ rotation around the $\varphi$-axis would be the same. However, as Fig.\ref{fig-xrd}b shows, the positions of (1 1 15) reflection ($\varphi = 145.8^\circ$) and (-1 -1 15) reflection ($\varphi = 32
3.5^\circ$) differ by $0.01^\circ$. It corresponds to the $0.005^\circ$ inclination of the $c$-axis towards elongated $a$-axis. Inclinations towards the other axes in the basal plane for the same sample are negligible. Thus, XRD clearly signifies reduced symmetry of the system.
In all studied single crystals distortion of both types was revealed. The parameters of the lattice distortion $\delta a/a$ have similar values in all studied crystals (see Table \ref{summarytable}).

To demonstrate that the discussed structural features originate from Sr intercalation, we show in Fig.\ref{fig-xrd}f for comparison 2$\theta /\omega$ curves for a bare Bi$_2$Se$_3$, grown under the similar growth conditions(temperatures, annealing and quenching regimes). The positions of the centers of two peaks coincide, the third peak slightly deviates, that correspond to elongation in one direction about 0.002\%, i.e. an order of magnitude smaller than in Sr-doped crystals. This observation proves that the anisotropic unit cell distortion is driven by the presence of the Sr atoms.

A puzzle is a scale of $H_{c2}$-anisotropy: how can $0.02$\% lattice deformation cause up to a factor of eight large and sample dependent $H_{c2}$ ratio?
 First, we should note that all XRD measurements in our paper as well as in all previous studies were performed at room temperature. Lowering the temperature suppresses the lattice oscillations making asymmetry more pronounced.  Low temperature XRD studies would be valuable. Another possibility is that the discovered small modification of the unit cell might be a consequence of stripe ordering of Sr atoms. Observation of such superstructures by transmission electron microscopy is hindered by high reactivity and mobility of the Sr under the electronic beam\cite{prmat}. Absence of correlations in Sr positions, seen by STM\cite{stm} might be simply due to a small analyzed area.

\subsection{Structural studies: block structure}

\begin{figure*}

\includegraphics[width=16cm]{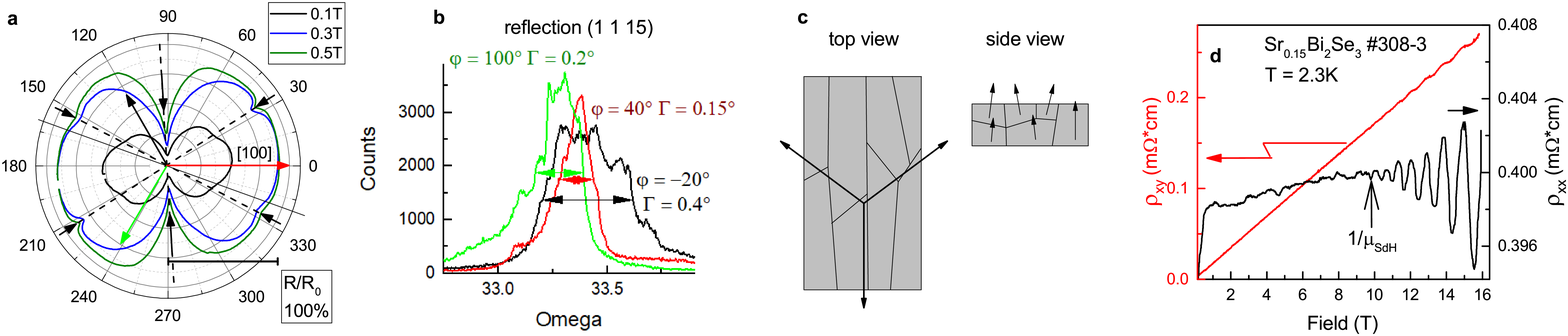}

\caption{Evidences for anisotropic block structure of the crystals. (a)AMR in sample Sr$_{0.15}$Bi$_2$Se$_3$ \#318-1 at $T=2.1K$ for various in-plane magnetic field values. (b)Rocking curves at reflection (1 1 15) for various $\varphi$ positions for sample Sr$_{0.15}$Bi$_2$Se$_3$ \#318-1. (c)Schematics of the block structure in the basal plane (left) and schematic side view of the same crystallities (right). Direction of a-axes in various blocks is the same, while c-axis direction vary from block to block, according to the rocking curves. (d)Resistivity (right axis) and Hall resistivity (left axis) versus magnetic field $\shortparallel c-$axis at 2K in Sr$_{0.15}$Bi$_2$Se$_3$ \#308-3 sample. The vertical arrow indicates an onset of the Shubnikov-de Haas oscillations.}

\label{fig-xrd-2}

\end{figure*}

Besides studies of the structural imperfections within one main block, we examined the angular distribution of the blocks within the crystal, i.e. we studied $\varphi$-dependence of the rocking curve. We have chosen 6-fold reflection (1 1 15) to probe anisotropy with higher resolution on $\varphi$ (each 60 $^\circ$). Fig.\ref{fig-xrd-2}a shows that rocking curve measured at $\varphi$ = 40$^\circ$ has smaller width than the $\pm 60^\circ$ neighbors of $\varphi$ = 40$^\circ$. As broadening of rocking curve is determined by misalignment of the blocks in the corresponding direction, the most realistic interpretation of these XRD data is a block structure shown in Fig.\ref{fig-xrd-2}b and Fig.\ref{fig-xrd-2}c. The direction of a-axis is the same for all
blocks (as the red curve is the narrowest). At the same time there is much spread of the c-axis direction from block to block (Fig.\ref{fig-xrd-2}c) that causes  widening of the rocking curve for $\varphi$ = -20$^\circ$.

Interestingly, the direction of blocks reasonably coincides with the direction of maximal elongation and maximal $H_{c2}$ in the same sample, as shown in Fig.\ref{fig-xrd-2}. The blocks are evidently separated by some transition regions (grain boundaries), indistinguishable by XRD.
To obtain a complete picture of the grain/boundary structure by real-space imaging of individual blocks additional studies by electron backscatter diffraction (EBSD)\cite{ebsd}, scanning X-ray nano-beam diffraction microscopy (SXRM)\cite{sxrm} and transmission electron microscopy(TEM) are needed.

Another indication of grain boundaries (suggested in Ref.\cite{kuntsevich}) is seen from the magnetotransport in perpendicular magnetic field. Figure \ref{fig-xrd-2}d shows magnetoresistance and Hall resistance at 2K for sample Sr$_{0.15}$Bi$_2$Se$_3$ \#308-3. Hall mobility $\mu_{Hall}\approx 400$ cm$^2/$Vs is straightforwardly obtained from the Hall slope to resistivity ratio. At the same time, Shubnikov-de Haas mobility is found from the field of magnetooscillations onset $\mu_{SdH}\sim 1/B_{ons}\approx 1000$cm$^2/$Vs. In single-component uniform system both mobilities are governed by the same scattering processes and $\mu_{SdH}/\mu_{Hall}$ is less than 2\cite{dassarma}. $\mu_{SdH}/\mu_{Hall}\sim 2.5$ ratio, observed in our case, as well as in Refs.\cite{jacs,srbises} implies that for some reason resistivity is too high. Indeed, if grain boundaries are responsible for this high resistivity, whereas low-disorder crystallites provide intensive magnetooscillations starting from relatively low fields, this high $\mu_{SdH}/\mu_{Hall}$ ratio is naturally explained.

\section{Statistics of anisotropy.}

If the nematicity were an intrinsic property of the SC condensate, then the discovered distortions would arrange the orientation of the nematic SC order parameter. In this case one would expect the anisotropy parameter(the ratio of $H_{c2}$ in the most and the less superconductive directions) to be weakly sample dependent, or correlated with critical temperature. In the opposite limit, the superconductivity should have 2-fold anisotropy because it couples strongly to the crystal 2-fold anisotropic structure.

\begin{figure}
\includegraphics[width=8cm]{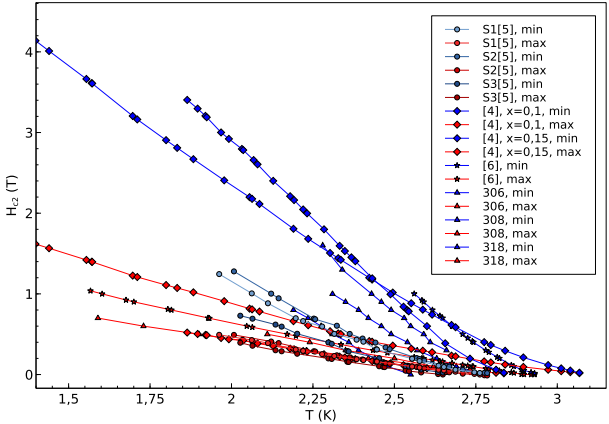}

\caption{Summary of the available data (our work and Refs. \cite{deVisser},\cite{chinese}, \cite{srbisenew})in-plane $H_{c2}$ values in the highest(shown in blue) and the lowest(shown in red) $H_{c2}$- directions in Sr$_x$Bi$_2$Se$_3$ single crystals. }
\label{comparison}

\end{figure}

Existence of structural anisotropy has already been indirectly indicated by extremely high, sample-dependent and theoretically unexpected\cite{fu2} in-plane superconductive anisotropy rates in Sr$_x$Bi$_2$Se$_3$\cite{srbisenew,deVisser,chinese} and Nb$_x$Bi$_2$Se$_3$\cite{shen}(in-plane ${H_{c2}}^{max}/{H_{c2}}^{min}$ ratio ranges from 2 to 8).

In Fig. \ref{comparison} we summarize the $H_{c}(T)$ data in both strongest and weakest superconductivity directions for all our samples and data from the other papers on Sr$_x$Bi$_2$Se$_3$. We observe no correlations between the anisotropy parameter and the critical temperature.

More detailed comparison of various observed SC parameters is made in Table \ref{summarytable}. Again, no correlation is seen between anisotropy factor, nominal Sr content, residual resistance ratio (RRR) and structural distortion.

\begin{table}
\caption{Summary of superconducting nematicity and structural parameters, observed by us and by the other investigators.}
\begin{tabular}{|c|c|c|c|c|c|}
  \hline
   Sample& x &$T_c$(K)&RRR&$H_{cmax}/H_{cmin}$&$\delta a/a$($10^{-4}$)\\
  \hline
  306& 0.1 &  2.73 &1.4&3&-1.93\\
  308-1 & 0.15 &  2.75 &1.5&6&8.5\\
  315 & 0.2 & 2.70  &1.7&4.5&2.65\\
  317 & 0.15 & 2.6&1.4 &4.8&2.1\\
  318 & 0.2 &  2.55 &1.5&8&2.17\\
  \cite{srbisenew} & 0.1 &2.92 &$\sim2$&4.5&-\\
  \cite{deVisser}& 0.1 &2.75 &1.75&6&\\
  \cite{deVisser}& 0.15 &2.9 &1.9&3&\\
  S1\cite{chinese,stm}&- &2.78&$\sim 1.5^*$&2.69&\\
  S2\cite{chinese,stm}&- &2.85 &$\sim 1.5^*$&$\sim$2.84&\\
  S3\cite{chinese,stm}&- &2.65&$\sim 1.5^*$&1.96&\\
  \hline
\end{tabular}
\label{summarytable}
\end{table}

This fact indicates, that some fine structural arrangement of Sr atoms is responsible for manifestation of the nematic superconductivity.
This conclusion is in line with recently reported strong correlations between crystal growth technology and superconductive properties\cite{prmat}.

\section{Discussion}
Structural anisotropy, discovered by us, absolutely does not contradict to the previous studies, even those reflecting the absence of such anisotropy , because we used much higher precision transport and XRD methods (see Appendix 3).

Anisotropic splitting of the sample into blocks naturally explains a complex character of the AMR above $T_c$: transport current might flow either along grain boundaries or across them, therefore AMR depends not only on current-to-field angle, but also on current-to-grain boundary angle. Non-monotonicity of magnetoresistance (Fig.\ref{fig-amr-2}c,d) might reflect distinct conductivity mechanisms in the grains and across the grain boundaries.

What are the reasons for the observed structural anisotropy? Our XRD studies reveal that one of crystallographical axes in the basal plane (let's call it $a$-axis) is typically co-aligned with the ampoule axis, i.e. with the vertical temperature gradient and, correspondingly, the growth direction.  It means that among all nucleated crystallites survive only those with $a$-axis orientation along the temperature gradient. They grow in huge blocks with different orientations of c-axis. During the subsequent cooldown inevitable stress occurs. In order to relax the stress, each huge block is split into smaller blocks with small c-axis misorientation.

Another source of stress is the internal pressure caused by introduction of Sr atoms into Bi$_2$Se$_3$ matrix, similarly to the Cu atoms\cite{intstress}. Internal pressure might be crucial for superconductivity like in FeSe\cite{fese}.


Possible scenario for the anisotropic SC may arise if the grain boundaries or some hidden defects host surface states with SC properties different from that of bulk carriers. Indeed, scanning tunneling microscopy (STM) studies of the Sr$_x$Bi$_2$Se$_3$ crystals with bulk $T_c$ about 3K revealed for the surface state in Sr$_x$Bi$_2$Se$_3$ SC gap closing at $T_c\sim 5$ K\cite{stm}, thus supporting the idea of surface SC. Interestingly, these STM studies revealed no in-plane anisotropy of SC properties in the basal plane. Thus, superconductivity along defects and grain boundaries could be a fruitful idea for explanation of the phenomenology of this system.

This scenario is however in contradiction with the large SC fraction ( up to 100\% in the best of our samples), detected in Sr$_x$Bi$_2$Se$_3$ also in Refs. \cite{chinese,jacs,shruti}, and with specific heat data, straightforwardly indicating bulk superconductivity both in Cu$_x$Bi$_2$Se$_3$\cite{HeatCap} and Sr$_x$Bi$_2$Se$_3$\cite{SrxBi2Se3Calor}. It is also inconsistent with sharp ($\Delta T \sim 0.1$K) transitions from normal resistance to zero. Indeed, interfaces of different blocks should be spread over properties and lead to smooth SC transition. Recent direct visualization of regular lattice of oval-shape Abrikosov vortices in Cu$_x$Bi$_2$Se$_3$\cite{andostm} also does not support surface-related scenario.

A promising scenario is  the recently suggested 2-component order parameter with p-wave pairing, belonging to D$_3$- symmetry class\cite{venderbos}. In this case the role of structural anisotropy is to favor the direction of nematic TSC orientation\cite{fu2}. This scenario seems to explain a bunch of experimental facts:
anisotropies of specific heat\cite{HeatCap}, magnetization\cite{asaba}, Knight shift\cite{NMR}, anisotropy survival under hydrostatic pressure \cite{pressure}, and proton bombardment \cite{radioactive}. The stability of the system originates from spin-orbit interaction \cite{michaeli}.
However it remains unclear why the value and the spread of anisotropy factor in Sr$_x$Bi$_2$Se$_3$ is so high?
 Indeed, within nematic topological theories, degree of anisotropy is governed by material-specific parameters and structural distortions only establish the  direction of the nematicity.

 Our research indicates a possibility , that small 0.02\% lattice elongation and $c$-axis incline might be consequence of much more anisotropic Sr atoms arrangement, causing much stronger modification of electronic spectrum, superconductivity, magnetoresistance. In case of ordinary s-wave superconductivity, the stability of the SC with respect to disorder is protected by the Anderson theorem. We believe, however, that in materials with such strong spin-orbit interaction, as doped bismuth chalcogenides, the role of structural distortion is more tricky. Indeed, in s-wave SC materials $H_{c2}$ can be limited either by spin mechanism or by orbital one. Spin-orbit interaction merges these mechanisms and in the presence of structural distortion makes the $H_{c2}$ value a complex matter.

All indications of the superconductive nematicity so far were performed in magnetic fields. There is no zero-field experiment (besides our room-temperature XRD measurements), demonstrating the anisotropy of the material, because transport search for anisotropy  without magnetic field is challenging. The resistive measurements in the vicinity of $H_{c2}$ probe essentially vortex phase, very sensitive to the pinning of the vortices.  Apparently, crystal anisotropy, preferable defect direction and preferable grain boundary direction make the pinning anisotropic. Crucial experiment to elucidate the role of boundaries would be an observation of SC anisotropy (or its absence) in $\mu$m scale and sub-$\mu$m scale samples. If superconductivity anisotropy in defect-free micro-scale samples would be absent, the overall idea of nematic SC in Sr$_x$Bi$_2$Se$_3$ becomes questionable.

Our results point to necessity of revision of a bunch of data to understand whether similar structural features correlate with anisotropy in Cu$_x$Bi$_2$Se$_3$ and Nb$_x$Bi$_2$Se$_3$. In respect of pinning, vortex phase for in-plane magnetic field configuration should be carefully examined in these materials.
Experiments on in-plane $H_{c1}$ anisotropy will allow to completely detune from vortices.

Thus our experimental research reveals the block structure and intrinsic structural anisotropy of superconducting Sr$_x$Bi$_2$Se$_3$. Now a question arises to what extent these properties are inherent to the other superconducting bismuth chalcogenides and whether the TSC scenarios in these materials should be revisited.

\section{Conclusion.}
Our results clearly demonstrate that: (i)two-fold structural anisotropy in Sr$_x$Bi$_2$Se$_3$ was observed and shown to be strong enough to both affect magnetotransport above $T_c$ and to be seen in XRD. (ii) The directions of the anisotropic features in transport correlate with the directions of the structural distortions. (iii) In addition to revealed unit cell distortions the crystals are anisotropically split into blocks, that points to the possible role of linear defects or grain boundaries in magnetotransport and vortex pinning. (iv) Despite  correlation between SC and structural anisotropies, the SC anisotropy parameter ($H_{c2}^{max}/H_{c2}^{min}$) is sample-specific.

All these items indicate that the nematicity in these materials is not a consequence of the topological superconductivity only and the relevance of structural factor should be taken into account.

\section{APPENDIX 1. Multi-block character of doped bismuth chalcogenides.}

We should note, that block structure of the superconducting bismuth chalcogenides is not just a property of our samples, but is rather inherent to these materials. Below we summarize numerous available indications:

\begin{enumerate}
\item{Except our paper, there is only one work (Ref. \cite{srbisenew})where detailed XRD studies are performed in high quality single crystals of Sr$_{0.1}$Bi$_2$Se$_3$.  The rocking curve, shown in Fig. 7b of Ref.\cite{srbisenew} clearly demonstrates several domains in accord with our observations (Fig.\ref{fig-xrd-2}b).}
\item{Recent direct STM observation of nematic superconductivity in Cu$_x$Bi$_2$Se$_3$\cite{andostm} also revealed at least two types of superconductive domains with different structural and superconducting properties and one type of non-superconductive domain all coexisting within the same crystal.}
\item{In the first observations of Shruti et al\cite{shruti}, powder XRD phase analysis in synthetised Sr$_x$Bi$_2$Se$_3$ revealed admixture of SrBi$_2$Se$4$ and BiSe phases. Similarly, admixtures of different fractions are seen in Nb$_x$Bi$_2$Se$_3$ \cite{kobayashiNb}. }
\item{In Ref.\cite{chinese} anisotropic sample-specific magnetoresistance deviates from 8-like shape, in agreement with our observations and multy-domain explanation. Magnetization in Ref.\cite{asaba} has also several features, explainable within multy-domain picture. }
\item{Normally the superconducting fraction in doped bismuth chalcogenides is not high\cite{kobayashiNb}, especially if the growth conditions are not properly adjusted\cite{prmat}. }
\end{enumerate}

\section{APPENDIX 2. Extremal multiblock sample 318-3.}

\begin{figure}[h]
	\includegraphics[width=9cm]{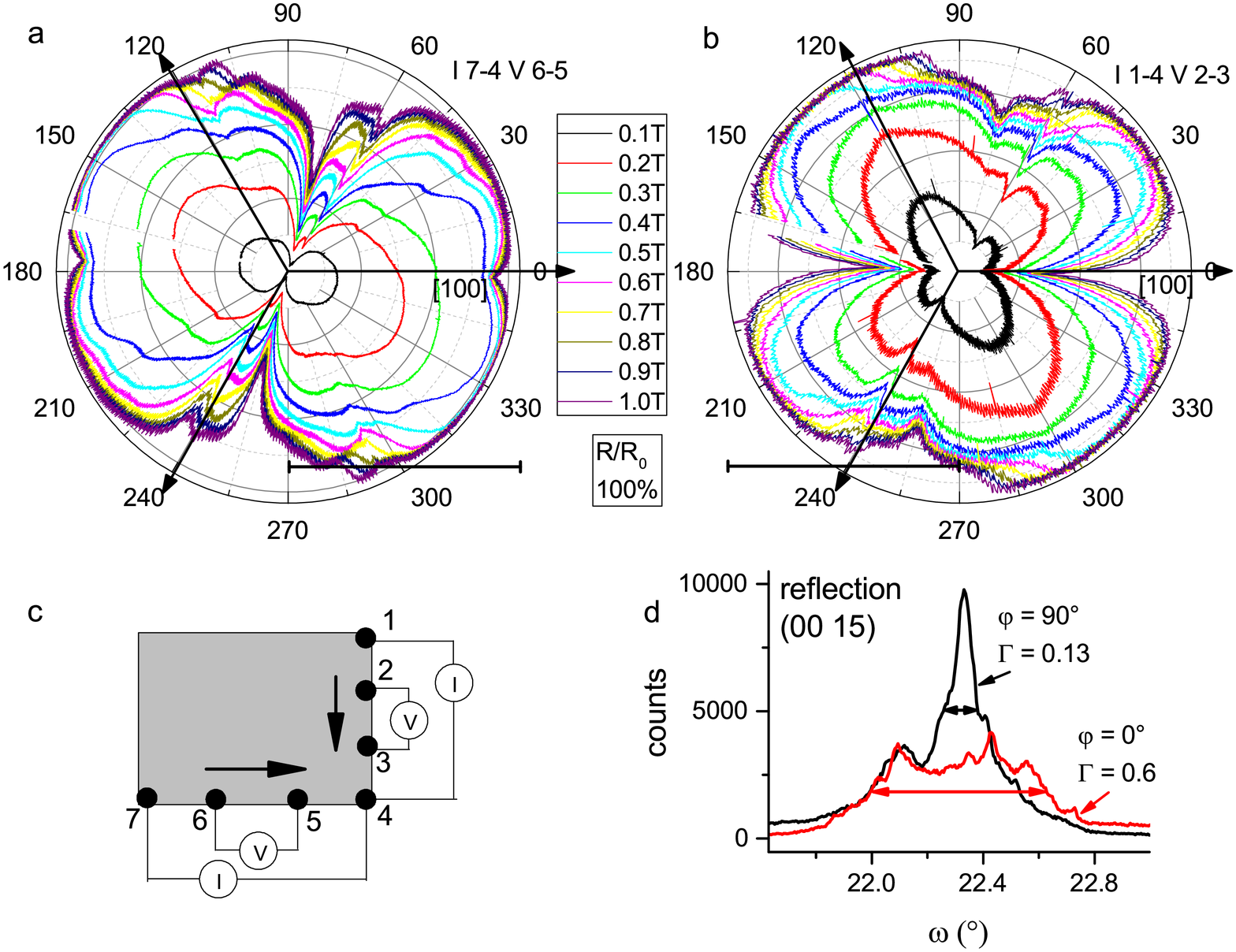}
	\caption{Anisotropic magnetoresistance at 2.2\,K for sample Sr$_{0.2}$Bi$_2$Se$_3$ \#318-3, (a) current 7-4, potential 6-5; (b) current 1-4, potential 2-3; (c) schematics of the contacts.}
	\label{fig2}

\end{figure}

  Fig.~\ref{fig2} shows the most spectacular magnetoresistance pattern for large multiblock sample Sr$_{0.2}$Bi$_2$Se$_3$ \#318-3 as an example of multidomain superconducting structure. In this figure one can see all complexity of the  possible mutual orientations of the blocks: (i) domains with almost perpendicular orientations (e.g. dip at 280$^\circ$ and dip at 180$^\circ$) and (ii) domains rotated by 60$^\circ$ (e.g. dip at 40$^{\circ}$ and dip at 100$^{\circ}$). Panels (a) and (b) compare magnetorsistance patterns for two current directions (schematics  of the current flow is given in the panel (c)). One can see that in panels (a) and (b) angular positions of superconducting dips are the same, but their amplitudes are different. This is natural, because the amplitude of these dips depends on path of the current, i.e. contributions of particular domains.

  Interestingly, the crystalline quality of this multidomain sample is rather high: it consists of crystallites with almost the same orientation, i.e. the rocking curves (shown in Fig. \ref{fig2}d) are rather narrow, below 0.6$^\circ$. Remarkably, the same rocking curves at (0 0 15) reflection clearly demonstrate the anisotropy in crystallite orientation, in agreement with Fig.\ref{fig-xrd-2}c: for orientation $\varphi=0^{\circ}$ the width is 0.6$^\circ$(red curve), while for perpendicular direction this width is 0.13$^\circ$(black curve). Since different blocks have different superconducting nematicity directions, our XRD data indicate  that superconducting nematicity direction is given not by orientation of parent crystalline axis but rather by some fine ordering of Sr atoms, that differs from block to block.

\section{Appendix 3. Comparison to Ref.\cite{srbisenew}.}

A rather similar recent research\cite{srbisenew} (high quality single crystal, single crystal XRD and anisotropic magnetotransport measurements) reported no visible structural distortions and no magnetotransport anisotropy above $T_c$ in Sr$_x$Bi$_2$Se$_3$, contrary to our results. We argue, that while the crystalline quality in Ref\cite{srbisenew} and our work is comparable , the XRD and transport measurement of Smylie et al,  did not resolve fine effects, reported by us.

In Ref.\cite{srbisenew} (see Fig. 7b )the measured full width at half maximum of the rocking curve is about 0.5$^\circ$, whereas in our the highest quality samples this value drops to 0.15$^\circ$, as shown in Fig.\ref{fig-xrd-2}. I.e the crystalline quality of our samples is at least not worse. Moreover, our results are confirmed on a number of single crystals with various Sr content.

Concerning the single crystal XRD studies, the research of Ref.\cite{srbisenew} in detail shows the absence of the improper peaks in diffraction patterns.
We carefully reproduced their measurements and, of course, obtained exactly the same: absence of $(hkl)$ reflections with $2h+k+l\neq 3n$, where $n$ is integer. This condition is however necessary, rather than sufficient for R${\overline{3}}$m crystallographic symmetry. Indeed, new peaks in diffraction pattern do not appear in response to slight structural distortion. On the contrary, they may arise, e.g. in thin films of Bi chalcogenides due to formation of twins\cite{twins}. In fact, the quality of our crystals and those in Ref.\cite{srbisenew} is so high, that there are no twins and only slight misorientation of blocks within the width of the rocking curve (less than 1$^{\circ}$). Therefore in order to reveal unit cell distortion and to identify symmetry breaking, one should analyze 120$^\circ$ and 240$^\circ$  $\varphi$-rotated asymmetrical reflections, as  we have shown in the body of the paper.

Two-fold anisotropy of AMR in the normal state was not observed in Ref. \cite{srbisenew}. It is rather expected, as all measurements were performed in a quite low (less than 1 T) magnetic  field. AMR in the normal state, discovered in our paper is negligible in that small fields. Fig. \ref{fig-amr-2} of our manuscript shows $\sim 1$\% anisotropy in magnetic field $\sim 10$ T.

\section{Appendix 4. Symmetry of the AMR.}
\begin{figure*}[t]
\includegraphics[width=16cm]{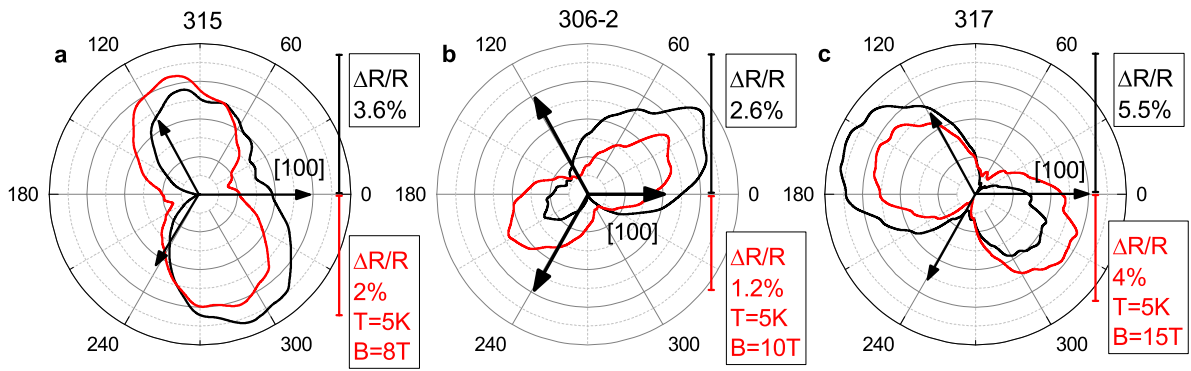}
\caption{ AMR for three representative samples above $T_c$ before (black curves) and after (red curves) subtraction of the first harmonic (panels a-c). Sample number, temperature, magnetic field and scale of the effect are indicated in the panels.}
\label{supp-fig-4}
\end{figure*}

In order to highlight the two-fold magnetoresistance symmetry (quite small effect above $T_c$), we subtracted the first harmonic $A\cos{\varphi} + B\sin{\varphi}$ from the AMR data. Figure \ref{supp-fig-4} shows the representative raw data and the data after subtraction. We did not found any correlation between the value and the direction of the first harmonic and the higher harmonics.
We believe therefore that the first harmonic comes from random admixture of the Hall effect to magnetoresistance. Indeed, the shape of the sample differs from the Hall bar and Hall component inevitably contributes to a voltage drop across the potential probes. There are two reasons for such admixture:
(i) unavoidable small sample inclination (up to 3$^\circ$) with respect to the plane of magnetic field vector rotation.
(ii) macroscopic sizes of the contacts, painted on the sides of the sample, that leads to partial current flow along $c$-axis.
Mechanism (i) allows to make an upper estimate of the first harmonic. In 10 T (typical field for our AMR measurements) and for 3$^\circ$ inclination, the perpendicular component is about 0.5T. From Fig.\ref{fig-xrd-2}d, assuming the lateral geometrical factor to be $\sim$1, we get the amplitude of the Hall effect to be about 2\% of $\rho_{xx}$. The experimentally observed first harmonic value in all cases was comparable or smaller.

To be fair, R$\bar{3}$m crystalline symmetry does not prohibit the first harmonic term in magnetoresistance, because magnetic field is an axial vector. However, charge carriers are located close to the $\Gamma$ point of the Brilluin zone and have rather isotropic in-plane properties. Therefore, there is no clear mechanism to cause such asymmetry in magnetoresistance. A search for such mechanisms (both theoretical and experimental) is a challenging forthcoming task.

Perpendicular $\sim 0.5$T component of magnetic field might serve as an additional mechanism to cause second harmonic $A\cos{2\varphi} +B\sin{2\varphi}$. We believe, however that this mechanism is irrelevant for the following reasons:
(i) in all our samples, magnetoresistance even in perpendicular field $\sim$10 T is rather weak, less than few \%  ($<0.5\%$ for sample 308,see Fig.\ref{fig-xrd-2}a, see also examples in Ref.\cite{jacs}). In 0.5 T field, the magnetoresistance is 40 times smaller, assuming that it is parabolic-in-$B$.
(ii) we systematically detect correlation between superconductivity direction and normal-state AMR direction. This correlation would be absent if randomly oriented perpendicular magnetic field admixture would be the case.

\section{Appendix 5. On the negative magnetoresistance.}

In Figs. \ref{fig-amr-2}e,f negative magnetoresistance (NMR) of the studied crystals in the parallel magnetic field weakly depends on temperature. Signatures of the negative magnetoresistance were also seen in Ref.\cite{deVisser} (Fig.3 blue curve and Supplementary materials, Fig. S7, panel a, black curve) and Ref.\cite{chinese} (Fig. 3 a,b,d,e where negative MR can be restored from the $R(T)$ data in various magnetic fields). The values and field scales of NMR are sample-specific, that points to the non-universal nature of the effect.

The mechanism of negative magnetoresistance is puzzling, and, we believe, may incorporate the following ingredients:
(i)quasi-two dimensional character of the in-plane transport and spectrum (discovered by Lahoud et al in Cu$_x$Bi$_2$Se$_3$\cite{lahoud} and indirectly seen in Sr$_x$Bi$_2$Se$_3$ from angular dependence of magnetooscillations\cite{jacs});
(ii) strong spin-orbit interaction, intrinsic to bismuth chalcogenides; (iii) presence of grain boundaries in the sample.

More exotic NMR mechanisms, like memory effects\cite{dmitriev} or Berry-phase induced magnetoresistance\cite{dai} could also be examined. We believe, however that increase of grain boundary transparency with magnetic field seems to be the most realistic and easily checkable candidate. Indeed, absence of NMR in the micro-size single-domain crystals would be a straightforward experiment to establish the role of grain boundaries. Such experiment is however yet to be done.

\section{Acknowledgments}

The work is supported by Russian Science Foundation (Grant N 17-12-01544).  The facilities of the LPI are used.
The authors thank A.L. Rakhmanov, R.S. Akzyanov, A.V. Sadakov, and Liang Fu  for discussions.

\end{document}